\documentclass[epj,final]{svjour}
\usepackage{wrapfig}

\usepackage{graphics}

\def\makeheadbox{{%
\hbox to0pt{\vbox{\baselineskip=10dd\hrule\hbox
to\hsize{\vrule\kern3pt\vbox{\kern3pt \hbox{Eur.Phys.J.B {\bf 54},
345-354 (2006)} \hbox{DOI: 10.1140/epjb/e2007-00006-x}
\kern3pt}\hfil\kern3pt\vrule}\hrule}%
\hss}}}

\begin{document}

\title{Stochastic Storage Models and Noise-Induced Phase Transitions}
\titlerunning{Stochastic Storage Models and Phase Transitions}
\author{{Serge Shpyrko}
\inst{1}$^,$\inst{2}$^,$\thanks{\email{serge\_shp\#yahoo.com}}
\and V.V.Ryazanov\inst{1}$^,$\thanks{\email{vryazan@kinr.kiev.ua}}}

\authorrunning{Serge Shpyrko \and V.V.Ryazanov}

\institute{Institute for Nuclear Research, Kiev, pr.Nauki, 47
Ukraine \and
Dept.of Optics, Palack\'y University,
T\v r. 17. listopadu 50, Olomouc, Czech Republic}
\date{Received 16 April 2006/Received in final form 26 October 2006 \\
Published online 13 January 2007}

%
\abstract{ The most frequently used in physical application
diffusive (based on the Fokker-Planck equation) model  leans upon
the assumption of small jumps of a macroscopic variable for each
given realization of the stochastic process. This imposes
restrictions on the description of the phase transition problem  where
the system is to overcome some finite potential barrier, or
systems with finite size where the fluctuations are comparable
with the size of a system. We suggest a complementary stochastic
description of physical systems based on the mathematical
stochastic storage model with basic notions of random input and
output into a system. It reproduces statistical distributions
typical for noise-induced phase transitions (e.g. Verhulst model)
for the simplest (up to linear) forms of the escape function.
We consider a generalization of the stochastic model based on the
series development of the kinetic potential. On the
contrast to Gaussian processes in which the development
in series over a small parameter characterizing the jump value is assumed
[Stratonovich R.L., Nonlinear Non\-equilibrium Ther\-mo\-dynamics,
Springer Series in Synergetics, vol.59, Springer Verlag, 1994],
we propose a series expansion directly suitable for storage models
and introduce the kinetic potential generalizing them.
} 
\maketitle
\section{Introduction}
\label{intro}


One of the aspects of modelling the behaviour of a complex physical
system consists in introducing a random process capable of
describing its essential properties. The most common (and in
practice almost unique) class of stochastic processes where
reliable results can be obtained is the class of Markov processes.
Said processes in their turn can be subdivided into different
families. The most widespread is the model of diffusion process
with Gaussian noise superimposing on the macroscopic dynamics. The
Poisson random processes (or "shot noise") present a second bench
point \cite{Horst} together with the former subclass covering the
most common physical situations. In the present paper we bring
into consideration the stochastic storage models based on
essentially non-Gaussian noise and treat them as a complementary
alternative to the diffusion approximation (that is to the
Gaussian white noise). We consider the phase transitions in such
models which resemble the noise-induced phase transitions
\cite{Horst}.

The class of stochastic storage models \cite{Prabhu,Brock}
presents a rather developed area of the stochastic theory. As
opposed to the common diffusion model \cite{Horst,Risken} it
contains as essential part such physical prerequisites: (i)
limitation of the positive semispace of states, (ii) the jumps
of a random physical process which need not be considered small;
(iii) essentially non-zero thermodynamic flux explicitely specified
by the process of random input.

The following material allows to conjecture that said models
provide more handy facilities of describing the noise-induced
phase transitions than diffusion ones \cite{Horst}. One of the
reasons in favour of this is the fact that typical probability
distributions there are not Gaussian but rather exponential and
gamma-distributions which are characteristic for, e.g., Tsallis
statistics. The approach of the present work is invoked to extend
the range of applicability of such models. Attempts have already
been made to apply them to the kinetics of aerosol coagulation
\cite{Ryazanov:1990}, to the problems of probabilistic safety
assessment methodology \cite{Ryazanov:1995} etc, and to relate
these processes to the Gibbs
statistics and general theory of dynamic systems \cite{Ryazanov:1993}.
One more possible application of the storage models consists in the
possibility of naturally introducing the concept of the lifetime
of a system
(the random time of existence of a given hierarchical
level) \cite{Prabhu,Ryazanov:1993,Ryazanov:2006,Chechkin:2005:1}.
It was shown \cite{Ryazanov:1993} that
the ambiguity of macroscopic behaviour of a complex system and the
existence of concurring evolution branches can be in principle
related to the finiteness or infiniteness of its average lifetime.

It is worth mentioning now that (at least) the simplest cases of
storage models do not require special probabilistic techniques,
and corresponding kinetic equations are treatable by means
of the Laplace (or Fourier) transform. Up to now such
models have not gained much recognition in physical problems. We
believe them to be rather promising especially in the approaches
based on modelling the kinetics of an open system, where the input
and release rates could be set from the physical background. In
the present work we do not intend to cover the variety of physical
situations akin to the storage models. Having discussed the form
of the stationary distributions for a set of input and release
functions (Sect.2) and their relation to noise-induced phase
transitions we reconsider the formalism of the kinetic potential
and fluctuation-dissipation relations (FDR) (Sect.3) and then pass
to the problem of reconstructing the underlying stochastic process
from the available macroscopic data (Sect.4). The material of
Sect.4 also considers the possibility of generalizing the
classical storage schemes to cover more realistic physical
situations. The concluding Section gives an example of an
application in the context of a practical problem of modelling a
nuclear fission process.

\section{Storage model as prototype to phase transition class models}
\label{sect:1}

Stochastic storage models (dam models) belong to a class of
models well known in the mathematical theory of random processes
\cite{Prabhu,Brock}. They bear a close relation to the queuing
theory and continous-time random walk (CTRW) schemes
\cite{Feller:2}. The visualization object for understanding
the physical ground of such a model is a reservoir (water pool),
the water supply to which is performed in a random fashion. The
random value $X(t)$ describing the bulk amount in a storage is
controlled by the stochastic equation:

\begin{equation}
X(t)=X(0)+A(t)-\int\limits_0^t r_{\chi}\left[X(u)\right] du\,.
\label{storage:eq}
\end{equation}
Here $A(t)$ is the (random) input function; $r(X)$ is the function
of output (release rate). Usually deterministic functions $r$ are
considered. In the simplest case it is constant:

\begin{equation}
r_{\chi}(X)=\left\{
\begin{array}{l}
a, X(t)>0  \\
0, X(t)=0 \,.
\end{array}
\right. \label{r:const}
\end{equation}
 The storage model (\ref{storage:eq}) is defined
over non-negative values $X \ge 0$, and the output from an empty
system is set to be zero (\ref{r:const}). Therefore the release
rate from (\ref{storage:eq}) is written as a discontinuous
function

\begin{equation}
r_{\chi}(q)\equiv r(q)-r(0+)\chi_q, \label{r:chi:1}
\end{equation}

\begin{equation}
\nonumber \chi_q = \left\{
\begin{array}{l}
1, \quad q=0 \,, \\
0, \quad q>0
\end{array}
\right.  \label{r:chi:2}
\end{equation}

More complicated input functions can be brought into
consideration. Analytical solutions are easy to find for escape
rates up to linear \cite{Prabhu,Brock}

\begin{equation}
r(X)=bX\, ; \quad \quad r(X)=a+bX \, . \label{r:linear}
\end{equation}

As to the random process $A(t)$ describing the input into the
system, it can be specified within various classes of processes.
For our purposes a partial case will be of special interest,
namely, that of L\'evy processes with independent increments
\cite{Prabhu,Feller:2}. It can be completely described by its
Laplace transform:

\begin{equation}
E(\exp(-\theta A(t))=\exp(-t\varphi(\theta))\, , \label{Levi}
\end{equation}
where $E(\dots)$ means averaging. The function $\varphi(\theta)$ is
expressed as

\begin{equation}
\varphi(\theta)=\int\limits_0^{\infty} \left( 1-\exp(-\theta x)
\right) \lambda b(x) dx   \label{phi}
\end{equation}
with
\begin{eqnarray}
\lambda=\varphi(\infty)< \infty \,; \quad \rho\equiv \lambda \int
x b(x) dx
= \varphi'(0)\, ; \nonumber  \\
 \mu^{-1}\equiv \int x b(x) dx =
\frac{\varphi'(0)}{\varphi(\infty)} \label{param} \,.
\end{eqnarray}
The function $b(x)$ and parameters (\ref{param}) have a
transparent physical meaning clear from the visualized water pool
picture of the model. Namely, $\lambda$ describes the intensity of
Poisson random jumps (time moments when there is some input into
the pool), and $b(x)$ is the distribution function (scaled to
unity) of the water amount per one jump with average value
$\mu^{-1}$. Thus, $\nu dx \equiv \lambda b(x) dx$ is the
probability dist\-rib\-ut\-ion of a generalized Poisson process
\cite{Feller:2} (for a ``pure'' Poisson process there were
$b(x)=\delta(x-\mu^{-1})$). For illustrative purposes the typical
choice
\begin{equation}
b(x)=\mu e^{-\mu x} \label{12}
\end{equation}
will be considered. In
this case the function (\ref{phi}) has the form

\begin{equation}
\varphi(\theta)= \frac{\lambda \theta}{\mu + \theta}
\label{phi:simple}
\end{equation}

The parameter $\rho$ gives the average rate of input into a
system representing thus an essentially non-zero thermodynamic flow. The
basic prop\-erty of the stochastic process under consideration is
thus the viol\-at\-ion of the detailed balance (absence of the
symmetry of the left- and rightwards jumps). This intrinsic
characteristics makes them a candidate for the systems
essentially deviating from the equilibrium (locating beyond the
"thermodynamic branch" \cite{Horst}).
 From this point of view the thermodynamic equilibrium
 of a storage model is achieved only in the degenerate
case $\lambda=0$, that is for a
system which occupies only the state $X=0$ (of course, the
equilibrium heat fluctuations are thus neglected).

Another property of the model consists in the finiteness of
jumps, on the contrast to the custom scheme of Gaussian Markov
processes with continuous trajectories. Therefore such models can
be believed to be more adequate in describing the systems with
fluctuations which can no more be considered small (for example,
systems of small size). We recover however the continuous-walk
scheme setting $\lambda \to \infty$, $\mu \to \infty$ and keeping
$\rho=\lambda/\mu$ finite. In this case the input is performed
with an infinite intensity of jumps of infinitely small size,
that is the system is driven by a Wiener-like noise process with
positive increments; if we limit
the release rate $r(X)$ with linear terms (\ref{r:linear}) the
process for the random variable then turns to that of
Ornstein-Uhlenbeck \cite{Risken,Vankampen,Gard}, and the storage
model presents its natural generalization. More specifically, one
can introduce the smallness parameter $\beta^{-1}$ [in equilibrium
situations with Gaussian noise it would equal to $kT$; generally
$\beta$ accounts for the environment and noise levels in a system
and can be related to the parameter in the stationary distribution
$\omega_{st}(X) \sim \exp(-\beta U_{\beta}(X))$] such that

\begin{equation}
\lambda_{\beta}=\lambda \beta \,, \quad b_{\beta}= \beta b (\beta
x) \, , \quad \varphi_{\beta}(\theta)=\beta
\varphi(\theta/\beta)\,, \label{stor:beta}
\end{equation}
from where the Gaussian case is recovered assuming $\beta \to
\infty$; the exponent in the characteristic function (\ref{Levi})
acquires now the form $\varphi(\theta)=\rho \theta +
\theta^2\sigma^2/(2\beta)$ of the Gaussian processes with drift.

It is instructive from the very outset to trace the relation of
the present models to the stochastic noise introduced by the
L\'{e}vy flights as well as to the processes encountered in the
CTRW. The non-Gaussian stable laws are described by means of the
characteristic function of their transition probabilities in the
form
\cite{Chechkin:2005:1,Feller:2,Metzler,Chechkin:2005:2,Chechkin:2003,Chechkin:2002,Sokolov:2003,Uchaikin,Zolotarev,Jespersen} \linebreak
$\exp(-t D |k|^\alpha)$ with the L\'{e}vy index $\alpha$, the case
$\alpha=2$ recovering the Gaussian law. The generalized central
limit theorem states that the sum of independent random variables
with equal distributions converges to a stable law with some value
of $\alpha$ depending on the asymptotic behaviour of the
individual probability distributions \cite{Feller:2,Uchaikin}.
In the case of the storage model the characteristic function
$\varphi(\theta)$ from (\ref{phi}) [where instead of $k$ there
enters $\theta$ after an appropriate analytical continuation in the
complex plane] comes in place of $D |k|^\alpha$. The finiteness of
$\varphi(\infty)\equiv \lambda$ indicates that the
trajectories of the storage process are discontinuous in time.
It is understandable that if the
functions $b(x)$ from (\ref{phi}-\ref{param}) have a finite
dispersion, the sum of many storage jumps will converge to the
Gaussian law with $\alpha=2$. From the physical picture of the dam
model, as well as from the analytical expressions like
(\ref{Levi}) we can see that the storage models present a class of
models where a mimic of the {\it long-range flights} is
effectively introduced, likewise in the L\'{e}vy flights, but the
nonlocality is achieved by virtue of the finiteness of allowed
jumps. Indeed, the trajectories of the centered process $A(t)-\rho
t$ present a saw-like lines with irregular distribution of
the jump sizes. Only in the limit of big times and scales it can
be viewed as a Wiener process with the variance $\langle
x^2\rangle = \sigma^2 t $ with $\sigma^2 \equiv \varphi''(0)$. On
shorter time scales the behaviour of the process models the
features akin to the superdiffusion (to the positive semiaxis),
and, on the contrary, a completely degenerated "subdiffusion" to
the left since the jumps to the left are forbidden. In this
context the function $\varphi(\theta)$ presents an {\it
effectively varying} L\'{e}vy index which ranges from the
superdiffusive region ($0 < \alpha < 2$) to negative meaning
suppression of the diffusion. The variable L\'{e}vy index for
diffusion processes is encountered in the models of
distributed-order fractional diffusion equations (see. e.g.
\cite{Chechkin:2002}). Actually, it is possible to bring into consideration
from the outset the input fuctions $b(x)$ pertaining to the ``basins of attraction''
of other stable distributions with L\'{e}vy indices $\alpha \neq 2$.
The storage schemes in which the functions $b(x)$ themselves are stable
L\'{e}vy distributions with power-like assymptotics $|x|^{-\alpha-1}$ are considered in \cite{Brock}.

The CTRW processes \cite{Feller:2,Metzler,Sokolov:2003}
are characterized by the joint distribution of the waiting times
and jumps of the variable.
The stochastic noise in the storage models is a narrow subclass
of CTRW where the waiting times and jump
distributions are {\it factorized}, and the waiting time
distribution is taken in a
single possible form  ensuring the Markov character of the
process \cite{Uchaikin}. This suggests a simple generalization to
a non-Markovian case. Namely,
assuming the storage input moments to be distributed  by an arbitrary law $q(t)$
instead of used $q(t)\sim \exp (-\lambda t)$ we arrive at generalized
CTRW schemes yielding semi-Markovian
processes which can be applied for introducing the memory effects
into a system.

The solution to the models (\ref{storage:eq}-\ref{Levi}) can be
found  either with the sophisticated apparatus of the mathematical
storage theory \cite{Prabhu,Brock} or directly by solving the
appropriate kinetic equation (see Sect.3,5). A considerable
simplification in the latter case is achieved in the Fourier space
where up-to linear release rates yield differential equations of
the first order (the situation is similar to the systems with
L\'{e}vy flights which are usually treated in the Fourier space;
note also the analogy to the method of the "Poisson representation" in the
chemical reactions problems \cite{Gard}).

 For the constant escape rate (\ref{r:const}) all
characteristics of the time evolution of the model are obtained in
the closed form \cite{Prabhu}. We mention just for reference for
$r_{\chi}=1$:

\begin{eqnarray}
\int\limits_0^{\infty}\exp(-s t) E\left(\exp(-\theta X(t))|X_0
\right) dt= \nonumber \\
\frac{[\exp(-\theta X_0)-\theta \exp(-X_0
\eta(s))/\eta(s)]}{[s-\theta+\varphi(\theta)]} \, , \nonumber
\end{eqnarray}
where $X_0=X(t=0)$, $\varphi(\theta)$ the same as in (\ref{phi})
and $\eta(s)$ satisfies a functional equation

\begin{equation}
\eta(s)=s+\varphi \left[\eta(s) \right] \, , \quad
\eta(\infty)=\infty \, .
\end{equation}

However the feature of interest now is the stationary
behaviour of the models of the class (\ref{storage:eq}). Even for
continuous functions $r$ and $b(x)$ the stationary distributions
$\omega_{st}(X)$ besides the continuous part $g(X)$ can have an
atom at zero, that is

\begin{equation}
\omega_{st}(X)=P_{0}\delta(X)+(1-P_0) g(X)\, , \label{10}
\end{equation}
where $g(X)$ is a probability distribution scaled to $1$, and
$P_{0}=lim_{t\to\infty}P(X(t)=0)$. The integral equation for
$g(X)$ from \cite{Brock} reads as:

\begin{equation}
r(X)g(X)=P_{0}\nu(X,\infty)+\int_{0}^{X}\nu(X-y,\infty)g(y)dy
\label{11}
\end{equation}
with the measure $\nu(x,\infty)\equiv
\lambda\int_{x}^{\infty}b(y)dy$. For the exponential shape of the
input function (\ref{12}) for which
$\nu(x,\infty)=\lambda\exp(-\mu x)$ the equation (\ref{11})  can
be solved for arbitrary release functions $r(X)>0$ ($C$ is found
from the normalization condition):
\begin{equation}
g(X) = \frac{C\exp\left(-\mu X+\lambda {\displaystyle \int}
\frac{\displaystyle \mathrm{d}X}{\displaystyle r(X)}\right)}{ \displaystyle r(X)} \,. \label{g:general}
\end{equation}
 The condition of the existence of the stationary
distributions for arbitrary input and release functions
from \cite{Brock} is the existence of some $w_0$ such that

\begin{equation}
\sup_{w\geq w_0}\int\limits_{y=0}^{\infty} \int\limits_{u=w}^{w+y} \frac{\mathrm{d}u}{r(u)}
\nu(\mathrm{d}y)
< 1   \label{cond}
\end{equation}
Similarly the expression for $P_0$
can be written for the general case \cite{Brock}. There is a
simple relation
$$ P_{0} =
\left[1+\int_{0}^{\infty}\langle\Gamma(x)\rangle\nu(\mathrm{d}x)\right]^{-1},
$$
between the weight of the zero atom $P_0$ and the average lifetime
$\langle\Gamma(x)\rangle$ (averaged random time of attaining the
zero level starting from a point $x$). The presence of the
non-zero $P_0$ indicates at the existence of idle periods where no
elements are present in a system. Such periods can be
characteristic for systems of the small size (in which the values
of fluctuations of a macrovariable are comparable to their
averages) \cite{Ryazanov:2006} and must influence essentially the
statistical properties of a system, for example, they impose
limitations on the maximal correlation time.

 The behaviour of the
models (\ref{storage:eq}) admits a pronounced property of
nonequilibrium phase transitions (change in the character of the
stationary distribution) which occur when one increases the value
of the average thermodynamic flow (parameter $\lambda$). The phase
transition points can be explored
by investigating the
 extrema of the stationary distribution  (cf.
the analysis of noise-induced phase transitions in \cite{Horst}),
that is for the case of (\ref{g:general}) -- from the condition
$\mu r(X)=\lambda-dr(X)/dX$. For
example, for the model with constant escape rate $r=a$ we get two
types of solutions: converging solution for small input rates and
the pool overflow (no stationary solution exists) if the average
input per time unit exceeds the output rate. The criterium for the
phase transition (\ref{cond}) in this case reads simply as
$\rho=a$. If $\rho>a$, no stationary distributions are possible.
For $\rho<a$ the stationary distribution
possesses additionally an atom $\delta(X)$ at $X=0$ with the
weight $P_0=1-\rho/a$. Explicitly for (\ref{12}) and $r=1$ the
stationary distribution $\omega_{st}(X)$ for $\rho\equiv \lambda/\mu<1$  is:

\begin{equation}
\omega_{st}(X)=P_0 \delta(X)+(1-P_0)(\mu-\lambda)
e^{-(\mu-\lambda)X} \, , \quad P_0=1-\rho \,.
\end{equation}

Consider now the exit function $r(X)=bX, b>0$ and the input in the form (\ref{12}).
This storage system does not have an atom at zero, and the
stationary probability distribution exists for all input rates -- there is
no overflow in the system:

\begin{equation}
\omega_{st}(X)=\mu^{\lambda/b}X^{\lambda/b-1}\exp(-\mu
X)/\Gamma(\lambda/b)  \label{omega:lin}
\end{equation}
($\Gamma(\lambda/b$) is gamma-function). The phase transition is
the modal change of the distribution function, which occurs at
$\lambda=b$, where the distribution changes its character from the
exponential $(\sim \exp(-X))$ to Gauss-like with a maximum at $X>0$
(Figure 1).
 This peculiarity of the stationary distribution can be
interpreted as a non-equilibrium phase transition induced by
external fluctuations. Such transitions are typical
\cite{Horst,Hongler} for the multiplicative type
of noise. They do not have their deterministic analogue and are
entirely conditioned by the external noise. The phase transition
at $\lambda=b$ manifests in the emerging of the nonzero maximum of
the distribution function although all momenta of the distribution
change continuously. As in the Verhulst model \cite{Horst} the phase
transition at $\lambda=b$ coincides with a point in which
$[D(X)]^{1/2}=E(X)$, where $D(X)$ is the dispersion, $E(X)$ is the
first moment of the distribution. With the choice of the input function
in the form
$b(x)=4 \mu^2 x \exp(-2\mu x)$ the stationary distribution is


\begin{eqnarray}
\omega_{st}(X) = \exp(-\lambda/b) (-\lambda/b)^{(1-\lambda/b)/2}
\times \nonumber
\\ X^{(\lambda/b-1)/2}
J_{\lambda/b-1}[2\sqrt{-\lambda X/b}] \nonumber
\end{eqnarray}
($J$ is Bessel function). The behaviour of the distribution is
qualitatively the same as on Fig.1 with the phase transition point
at $\lambda=b$ as well. In both cases we have the phase transitions
which are caused by an {\it additive} noise (which does not depend on
the system variable). The existence of phase transitions for the
additive noise is closely related to the long-range
character of the distribution function of the noise and such
transitions were discovered in systems with, e.g. L\'{e}vy type of
additive noise \cite{Chechkin:2005:1,Chechkin:2005:2} where the
structural noise-induced phase transitions are
conditioned by the trade-off between the long-range character of
the flights and the relaxation processes in the model. Analogous conclusions
for other types of superdiffusive noises are also drawn in
\cite{Jespersen,Hongler} etc. We can thus state that the effective
long-rangeness in the storage models leads to similar effects
causing the modal changes of the distribution function which can
be interpreted as a nonequilibrium noise-induced phase transition.

\begin{figure}[th]
\resizebox{0.55\textwidth}{!}{%
  \includegraphics{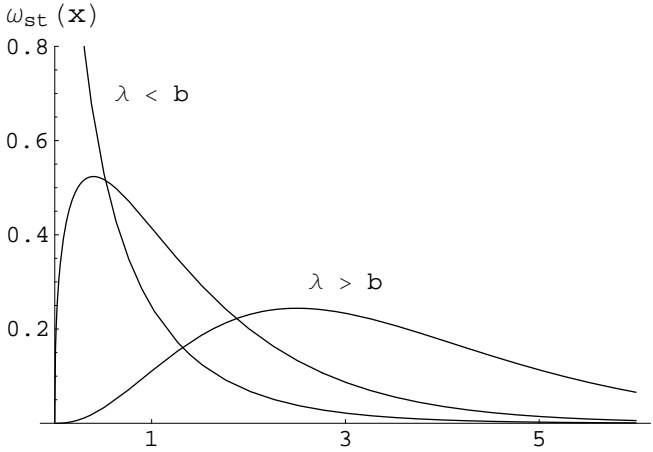}
} \caption{Stationary distribution function (\ref{omega:lin}) in
the storage model with $r(X)=bX$. Phase transition with increasing
input intensity $\lambda$.} \label{fig1}
\end{figure}

More complicated example is the rate function $r(X)=a+bX$ and the
input rate (\ref{12}). In this case
$$
 \omega_{st}(X) = P_{0}\left[\delta (X) +\frac{\lambda}{b}
a^{-\lambda /b} \exp\{-\mu X\}(a + bX)^{\lambda /b - 1}\right] \,,
$$
 $$P_{0}^{-1}=1+(\mu a/b)^{-\lambda
/b}(\lambda/b)\exp\{\mu a/b\}\Gamma(\lambda/b; \mu a/b)$$
($\Gamma(x ;y)$ is incomplete gamma-function). It combines two
previously considered models: there is an atom at $X=0$ with
weight $P_0$ and there is the phase transition at critical
$\lambda$ where the distribution switches from exponential to
Gaussian-like. This critical value is $\lambda_{cr}=a\mu+b$. If
$b\to 0$, it coincides with that for
the model $r=a$, and if $a\to 0$ - with the results of $r=bX$
model (\ref{omega:lin}).

For a realistic release function $r(X)=bX-cX^{2}(1-X)$, ($c,b\geq
0$) corresponding to, for example, the nonlinear voltage-current
characteristics, the solution to (\ref{11}) for exponential input
(\ref{12}) (expression (\ref{g:general})) yields [in this case, like
for the linear model $r=bX$, $P_0=0$]:

\begin{eqnarray}
g(X)=\frac{N \exp(-\mu X) (X+c)^{\lambda/b}}{(bX-cX^{2}+cX^{3})
\mid cX^{2}-cX+b \mid ^{\lambda/2b}}\times \nonumber
\\
\times
 \exp\left(\frac{\lambda c}{2 b \sqrt{c \mid c-4b \mid}}
\arctan\frac{c(2X-1)}{\sqrt{c \mid c-4b\mid}}\right), c<4b 
 \nonumber
\end{eqnarray}
(we consider the case $c <4b$ only because natural restrictions on
the drift coefficient impose $r(X)>0$ \cite{Brock}). The phase
transition points can be explored from
 the custom analysis of the value of the corresponding cubic determinant $Q$.
The number of phase transitions varies from one ($Q>0$) to three
at $Q<0$ depending on the relations between $\mu^{-1}$, $b/c$, and
$\lambda/b$. This latter example can be compared to the
nonequilibrium regimes found in the quartic potential well driven
by an additive L\'{e}vy-type noise
\cite{Chechkin:2005:2,Chechkin:2003}; as in the above systems the
additional criticality is achieved due to a long-range character of the
additive noise.

The distribution functions and phase transitions in this class of
models are common to various physical systems. Relative simplicity
in mathematical treatment allows us to propose them as a handy tool
for modelling physical phenomena of stochastic nature. We will
show that this class of models based on the Poisson noise presents
a prototype for stochastic modelling complementary to the commonly
considered Gaussian noise. This latter, as the simplest case of
phenomenologically introduced stochasticity, became in fact the
most recognized way of introducing noise, and various enhancements of
stochastic description meant merely an extension of the Langevin
source of a Gaussian nature introducing the small parameter of jumps
value (see later in the Section 4). The use of the Poisson noise,
starting from the storage model as its basic case, allows an
extension to more complicated cases as some regular development
series in a small parameter
as well. In contrast to the Gaussian
scheme, it does not require at the very beginning the smallness of
jumps thus it is able of describing more adequately a wide class
of physical phenomena where this assumption does not have its
physical justification. As an example we mentioned a thermodynamic
system of a small size, where the random value is the number of
particles. The Gaussian assumptions are valid only for large size
(in comparison to the rates of input or output) of a system, when
the diffusion approximation can be used. The smaller is the
system, the worse is the description in terms of minor random
jumps (basic for Gaussian scheme), and, vice versa, more reliable
becomes the description based on the Poisson character of a random
process.

\section{Fluctuation-dissipation relations and formalism of the kinetic
potential} \label{sect:2}

This section is a brief reminder of the formalism of the kinetic
potential \cite{Stratonovich} which is appropriate for presenting
the properties of a Markovian random process in a compact form.

The primary concept of the Markov process is the transition
probability for a random value $B$ to jump from the initial state
$B_1$ at the time moment $t_1$ into state $B_2$ at $t_2$, that is
the probability $\omega_{21}\equiv\omega(B_2,t_2| B_1, t_1)$.
 The idea of Markovian behaviour imposes obvious restriction on the values
$\omega_{21}$, so that they possess a superposition property
$\int\omega_{32}(B|B')\omega_{21}(B'|B'')\mathrm{d}B'=\omega_{31}(B|B'')$,
$(t_1<t_2<t_3)$; in other words, form a continuous semigroup in
time (the reverse element of this group for dissipative processes
is not defined), so that all characteristics of a system can be derived
from the infinitesimal generators of the group. In normal
language, we bring into consideration the probabilities per time
unit $(1/\tau)\omega(B+\Delta,t+\tau|B, t)$ ($\tau \to 0$). To
characterize this function it is useful to consider its
moments which are called ``kinetic coefficients'':

\begin{equation}
K_n(B,t)\equiv \lim_{\tau\to 0}\frac{1}{\tau}\int
\omega(B+\Delta,t+\tau|B,t)\Delta^n \mathrm{d}\Delta \,. \label{K}
\end{equation}

The stationarity of the Markov process assumes that $K_n$ are
time-independent. The kinetic equation for the distribution
function of the process reads as
\cite{Risken,Vankampen,Stratonovich}

\begin{eqnarray}
\frac{\partial \omega(B,t)}{\partial
t}=\sum\limits_{n=1}^{\infty}\left( - \frac{\partial^n }{\partial
B^n}\right)\frac{K_n(B)}{n!}\omega(B,t)\,;
\label{kin:equat} \\
\omega(B,t)=\int \omega_{tt_1}(B|B')\omega_{t_1}(B')\mathrm{d}B'\,.
\nonumber
\end{eqnarray}

The kinetic potential is defined as the generating function of the
kinetic coefficients \cite{Stratonovich}:

\begin{equation}
V(-\theta, B)\equiv \sum\limits_{n=1}^{\infty}
K_n(B)\frac{(-\theta)^n}{n!}\,. \label{kin:pot}
\end{equation}

Thus the kinetic coefficients can be expressed as

\begin{equation}
K_n(B)=\frac{\partial^n}{\partial
(-\theta)^n}V(-\theta,B)_{|\theta=0}
\end{equation}
for $n=1,2,\dots $. With (\ref{kin:pot}) the equation
(\ref{kin:equat}) can be written compactly:

\begin{equation}
\frac{\partial \omega(B,t)}{\partial t}=\mathcal{N}_{\partial, B}
V\left(-\frac{\partial}{\partial B},B \right) \omega(B,t) \, .
\label{kin:equat:pot}
\end{equation}
In (\ref{kin:equat:pot}) the notation ${\cal{N}}_{\partial,B}$ means the
order of the differentiation operations: they should follow all
actions with the multiplication by $K_n$ as it is seen from
(\ref{kin:equat}).

An example of the kinetic potential for the simplest and most
utilized stochastic process is

\begin{equation}
V(-\theta,B)=K_1(B)(-\theta) + \frac{1}{2}K_2(B)\theta^2 \,.
\label{kin:pot:gauss}
\end{equation}
With the choice $K_1(B)=-b\cdot B$, $K_2(B)=D=\mathrm{const}$,
$K_{n>2}\equiv 0$ the corresponding kinetic equation is then the
Fokker-Planck equation for the Orstein-Uhlenbeck process
\cite{Risken,Vankampen} which describes a system with linear relaxation
towards the stationary solution in the Gaussian form
$$\omega_{st}(B)\sim \exp\left(-\frac{b B^2}{D}\right)\,.$$
 Note that the
kinetic potential for a process driven by a L\'{e}vy flight noise
has thus the generic form
$V(-\theta,B)=-\theta K_1(B) + D\theta^\alpha$ and assumes the
kinetic equation of the formally fractional order which is not
reduced to the series in (\ref{kin:equat}); one uses instead its
plausible generalization which can be encountered elsewhere (e.g.,
\cite{Metzler,Sokolov:2003,Uchaikin,Zolotarev,Jespersen} etc).

As another example we write down the form of the kinetic potential
for the class of storage models of Section 2. The kinetic
potential $V(-\theta,B)$ through the transition probabilities of a
Markov process is written as

\begin{eqnarray}
V(-\theta,B,t)=\lim_{\tau \to {0}}\frac{1}{\tau}\left[E\left(e^{-\theta(X(t+\tau)-X(t))}\mid
X(t)\right)-1\right]\,; \nonumber
  \\
E\left(e^{-\theta(X_{3}-X_{2})}\mid X_{2}\right)=
\int e^{-\theta(X_{3}-X_{2})}\omega_{t_{3}t_{2}}(X_{3}\mid X_{2})\mathrm{d}X_{3}
\nonumber
\\
= E(e^{-\theta X_{3}}\mid
X_{2})E(e^{\theta X_{2}})\,,  \hspace{2em}   \label{stoch:pot:der}
\end{eqnarray}
where $X_{k}=X(t_{k})$. Inserting there the Laplace transform of
the random value from (\ref{storage:eq}), we obtain
\cite{Ryazanov:1993}, for an elementary derivation see Appendix:

\begin{equation}
V(-\theta,B)=-\varphi(\theta)+\theta r_{\chi}\left( B \right) \, ,
\label{stoch:pot}
\end{equation}
where $\varphi$ is defined in (\ref{phi}), and $r_{\chi}(B)$ in
(\ref{r:chi:1},\ref{r:chi:2}).

It is handy to introduce another generating function called the
``image'' of the kinetic potential \cite{Stratonovich}. Namely,
let $\omega_{st}(B)$ be the stationary solution $\dot{\omega}=0$
of the kinetic equation (\ref{kin:equat}). The image of the
kinetic potential $V$ is defined as

\begin{equation}
R(y,x)\equiv {\displaystyle \frac{\int \exp(x B) \omega_{st}(B) V(y,B) \mathrm{d}B}{\int
\exp(x B) \omega_{st}(B) \mathrm{d}B} } \label{R:1}
\end{equation}
or, in the notation of the transition probabilities,

\begin{eqnarray}
R(y,x)\equiv \lim_{\tau\to 0}\frac{1}{\tau}\times\int\int
\mathrm{d}B_1 \mathrm{d}B_2 \exp(x B_1)
\label{R:2} \\
\times\frac{\big[ \exp(y(B_2-B_1))-1\big]  \omega(B_2,
t+\tau|B_1,t)\omega_{st}(B_1) }{\int \exp(x B) \omega_{st}(B)
\mathrm{d}B}\,. \nonumber
\end{eqnarray}
The series of $R(y,x)$ over $y$:
\begin{equation}
R(y,x)=\sum\limits_{n=1}^{\infty} \kappa_n(x)\frac{y^n}{n!}
\label{R:kappa}
\end{equation}
defines new coefficients $\kappa_n(x)$ being the image of $K_n$:

\begin{equation}
\kappa_n(x)=\frac{\int K_n(B)\omega_{st}(B)\exp(x B) \mathrm{d}B}{\int
\omega_{st}(B)\exp(x B) \mathrm{d}B}\,.\label{kappa}
\end{equation}
We note by passing that the variable $x$ in (\ref{R:1}) or
(\ref{kappa}), presenting merely a variable over which the Laplace
transform of the process variables is performed, can be also
understood as a (fictive or real) thermodynamic force.  This
interpretation is clarified when we look at
the ``pseudo-distribution'' $\exp(x B) \omega_{st}(B)$ where $x B$
stands for an amendment to the free energy of a system
\cite{Stratonovich}.

The reconstruction of a stochastic random process assumes that
knowing macroscopic information about system we make
plausible assumptions as to the fluctuating terms of the kinetic
equation, that is we try to construct the matrix of the transition
probabilities $\omega_{ij}$ in any of equivalent representations
(\ref{K}), (\ref{kin:pot}) or (\ref{R:1}) leaning upon some
macroscopic information about the random process. As the latter,
we can understand the following two objects: 1) the
stationary distribution $\omega_{st}(B)$, and 2) ``macroscopic''
equations of motion which are usually identified with the time
evolution of the first momenta of $\omega(B,t)$ and hence the
kinetic coefficient $K_1(B)$ (in the case of the sharp probability
distribution where one can identify the "macroscopic variable" at all).
For example, for the storage model scheme the
problem is inverse to that considered in Section 2: knowing the
macroscopic relaxation law and the shape of the stationary
distribution  we then try to reconstruct the input function
$\varphi(\theta)$. The relaxation law is given by the balance of
the averaged input and release rate $\rho - r(X)$ (in the present
class of storage models the input rate is $X$-independent, the
generalization is considered further). Then, given $r(X)$
 one can set into correspondence to it the input
function $\varphi(\theta)$ yielding a given distribution $\omega_{st}(X)$:

\begin{equation}
\varphi(\theta) = \theta\frac{\int r(X)\omega_{st}(X)\exp(-\theta X)\mathrm{d}X}
{\int \omega_{st}(X)\exp(-\theta X)\mathrm{d}X}\,.
\label{phi:reconst}
\end{equation}

The relations between said objects and the remaining part of
the stochastic information contained in the process are called
fluctuation-dissipation relations (FDR). These relations express
the property of time reversibility of the transition probabilities
(detailed balance). In the representation of the image of kinetic
potential (\ref{R:1}), (\ref{R:2}) they are written in the most
elegant fashion \cite{Stratonovich}:

\begin{equation}
R(y+x,x)=R(-\varepsilon y, \varepsilon x) \, , \label{FDR:1}
\end{equation}
where $\varepsilon=\pm 1$ according to the parity of the variable.
The particular case of the FDR in the form (\ref{FDR:1}) at $y=0$
represents the ``stationary'' FDR

\begin{equation}
R(x,x)=0  \, . \label{FDR:2}
\end{equation}
The FDR in the form (\ref{FDR:2}) hold for any system in the
stationary state with no assumption about the detailed balance,
that is the system needs not to be in the equilibrium state.
Indeed, it is easy to check that (\ref{FDR:2}) is just an another notation
of the equation for the stationary distribution
$ {\mathcal N}_{\partial,B}V\left(-\partial/\partial
B,B\right)\omega_{st}(B)=0$ (see Appendix).

\section{Reconstruction of the random process}
\label{sect:3}

The problem of reconstructing a random process in the notations of
the preceding section is formulated as a set of algebraic
equations for a function $R(y,x)$. Thus, given functions
$\kappa_1(x)$ and $\omega_{st}(B)$ we must find $R(y,x)$ which
identically satisfies the relation (\ref{FDR:1}) (for the system
in equilibrium) or (\ref{FDR:2}) (for the stationary system with
no assumption about the thermal equilibrium and detailed balance).
The problem, of course, has many solutions since those conditions
do not define the function $R$ uniquely. There exists the
``FDR-indeterminable information'' which hence should be borrowed
from some additional criteria to be imposed on the equations in
order to close the problem, which means confining ourselves within
some class of the stochastic processes. The
kinetic potential representation allows us to elucidate clearly the nature of the
approximations made.

\subsection{``Gaussian'' scheme}
\label{subs:1}

The standard reconstruction procedure considers the possibility of
setting the ``FDR-indeterminable'' functions negligibly small,
that is introducing a small parameter over which the kinetic
potential can be developed in series \cite{Vankampen,Stratonovich}.
Thus
the generalization of the ``bare Gaussian'' model showed
in the example above (\ref{kin:pot:gauss}) is achieved. In
the series

$$ V(y,B)= K_1 y + K_2 \frac{y^2}{2!} +K_3 \frac{y^3}{3!} + \dots $$
successive coefficients $K_n$ decrease progressively as
$\beta^{1-n}$. This relation can be expressed introducing the
family of kinetic potentials labelled by the large parameter
$\beta$ \cite{Stratonovich} (compare with (\ref{stor:beta}))

\begin{equation}
V_{\beta}(-\theta,B)\equiv \beta V (-\theta/\beta, B)\,,
\label{V}
\end{equation}

 This is a common approximation for a random process
consisting in the fact that its jumps are small. If we keep only
two terms $K_1$ and $K_2$, the second coefficient for one variable
can be restored from FDR exactly. As an example, we write the
kinetic potential reconstructed up to 4-th order (from the formula
(\ref{FDR:1}) applying development in the powers of $y$):

$$
R_g(y,x)=y \kappa_1(x)
\left(1-\frac{y}{x}\right)+y^2(y-x)^2\frac{\kappa_4(x)}{4!} \, ,$$
where the coefficient $\kappa_4(x)$ is arbitrary (indeterminable
from FDR). The first term describes the base variant corresponding
to (\ref{kin:pot:gauss}).

\subsection{``Storage'' scheme}
\label{subs:2}

The assumption of small jumps leads to the possibility to neglect
the higher order kinetic coefficients $K_n$, constructing a
stochastic process by the ``Gaussian'' scheme (G-scheme). We
suggest an alternative approach which can be regarded as
complementary to the G-scheme and does not require the assumption
of the small jumps. Like G-scheme, it has its basic variant which
is well treatable mathematically.

We assume now that the kinetic coefficients $K_n(B)$ are
expandable into series over the variable $B$:

\begin{equation}
K_n(B)=k_{n,0}+ k_{n,1}B + k_{n,2}\frac{B^2}{2}+ \dots
k_{n,l}\frac{B^l}{l!}\dots \,.  \label{S:series}
\end{equation}

Possibility of truncating these series implies that
$k_{n,l}$ in (\ref{S:series}) contain a small parameter $\gamma$
which decreases them progressively with the growth of the number
$l$: $k_{n,l}\sim \gamma^l$ for $n=2,3,4,...$. In the coefficient
$K_1(B)$ determining the macroscopic evolution we however keep the
macroscopic part $r_{\chi}(B)$ whose development on $B$ does not
depend on $\gamma$: $K_1(B)=-r_{\chi}(B) + \sum k_{1,l}B^l/l!$.

The image of the kinetic potential thus turns out to be a
development into series

\begin{equation}
R(y,x)= -y\overline{r_{\chi}(x)} - \sum\limits_{l=0}^{\infty}
\frac{1}{l!} \langle A^l(x) \rangle \varphi_l(y)\,,
\label{S:image}
\end{equation}
here
$$\overline{r_{\chi}(x)}=\frac{\int r_{\chi}(B)
\omega_{st}(B) \exp (xB)\mathrm{d}B}{\int \omega_{st}(B)
\exp (xB)\mathrm{d}B} \,,$$

$$
\langle A^l(x)\rangle \equiv \frac{ \int B^l\omega_{st}(B)\exp(x
B) \mathrm{d}B}{\int \omega_{st}(B)\exp(x B) \mathrm{d}B } \, ,
$$
and $\varphi_l$ are defined through coefficients in
(\ref{S:series}):

$$
-\varphi_l(y) \equiv \sum\limits_{n=1}^{\infty}
y^n\frac{k_{n,l}}{n!} \, , \quad l=0,1,2,\dots
$$

The series (\ref{S:image}) can be considered as a development on
the base $\{1, \langle A(x)\rangle, \langle A^2(x)\rangle, \dots
\}$ which is a natural base of the problem following from the
peculiarities of its stationary distribution. The coefficient
$\varphi_0(y)$ at $y=-\theta$ has the same meaning as the function
$\varphi$ from (\ref{phi}) and truncating (\ref{S:image}) up to it
reproduces the storage scheme of Sec.2. This is a generalization referring
to the multiplicative noise processes $dX(t)= -r_{\chi}(X)dt +
dA(t;B)$ instead of (\ref{storage:eq}) with the characteristics
function of the noise $E(\exp(-\theta A(t)))=\exp(-t[\varphi_0 + B
\varphi_1 + \dots])$ instead of (\ref{Levi}).

From the equation $R(x,x)=0$ applying it to (\ref{S:image}) we
obtain the reconstructed scheme for a random process (``S-scheme''):

\begin{eqnarray}
R_s(y,x) = -\sum\limits_{l=0}^{\infty}\frac{1}{l!}
\langle A^l(x) \rangle \left[ \varphi_l(y) - \frac{y}{x}\varphi_l(x)\right] = \hspace{2em}
\label{eq2} \\
-y\left(\overline{r_{\chi}(x)}- \overline{r_{\chi}(y)}
\right) -
 \sum\limits_{l=1}^{\infty}\frac{1}{l!} \varphi_l(y)
\big(\langle A^l(x) \rangle- \langle A^l(y) \rangle\big) \nonumber \,.
\end{eqnarray}

Keeping in mind that the coefficient $\kappa_1(x)$ is supposed to
be known the last expression can be rewritten as

\begin{eqnarray}
R_s(y,x)=y\kappa_1(x)\left( 1-\frac{\kappa_1(y)}{\kappa_1(x)}
\right) - \nonumber
\\
\sum\limits_{l=1}^{\infty}
\frac{1}{l!}\eta_l(y) \left(\langle A^l(x)\rangle - \langle A^l(y)
\rangle \right) \label{S:potent}
\end{eqnarray}

The coefficients $\eta_l(y)$ at $l=1,2,\dots$ are
dissipative-indeterminable. They are given by

$$\eta_l(y)=
-\sum_{n=2}^{\infty} y^n k_{n,l}/n! \equiv \varphi_l(y) -y
\varphi_l'(0) \,.
$$

If we set $\varphi_l=0$ for $l=1,2,\dots$ in (\ref{S:image}-\ref{S:potent}), we
recover the kinetic potential of the ordinary storage model
(\ref{stoch:pot}) which can be restored from the stationary
distribution exactly. The above formula (\ref{phi:reconst}) is a particular
case of the application of this scheme.

\section{Conclusion. An example of possible application}
\label{sect:4}

Both schemes
sketched above - that is the common G-scheme and suggested
complementary S-scheme of the stoch\-ast\-ic reconstruction both lean
upon two basic stochastic models which use respectively the
assumption of Gaussian and Poissonian nature of the random noise.
They both apply the series development of the kinetic potential on
a small parameter. Keeping infinite series leads to
identical results in both cases. However, in real physical
problems we use to truncate the expansion series keeping small
finite number of terms. According to the physical situation and to
the nature of the random noise either one, or another scheme would
give a more reliable convergence.

Consider now an example of the storage model with the linear release
rate $r=bX$ (\ref{r:linear}) and generalized input which is now
$X$-dependent (Sect.4.2). Find the solution of the simplest linear
dependence of the input function on $X$ with the kinetic potential
(\ref{kin:pot}) set as
\begin{equation}
V(-\theta,X) = - \varphi_0(\theta)  + \theta b X -c X
\varphi_1(\theta)\,. \label{V:gen:exam}
\end{equation}
 The first two terms in (\ref{V:gen:exam})
describe the usual storage model with linear release, and the last
term is the amendment to the input function proportional to $X$
(cf. (\ref{S:series}-\ref{S:image})). The parameter $c$ controls the intensity
of this additional input. The equation for the Laplace transform
$F(\theta)\equiv E(\exp(-\theta X))$ of the stationary
distribution is $V(-\theta, X\to d/d\theta)F(\theta)=0$, where
the differentiation refers only to the function $F$. Its solution for
$V$ in (\ref{V:gen:exam}) is

\begin{equation}
-\log F(\theta) = \int\limits_0^{\theta} \frac{\varphi_0(u)du}{bu
- c \varphi_1(u)}
\end{equation}

For illustration specify now the input functions as

\begin{equation}
\varphi_0(\theta)= \varphi_1(\theta) = \frac{\lambda \theta}{\mu +
\theta} \, ,
\end{equation}
which correspond to the exponential distribution functions of input jumps
(\ref{phi},\ref{12}) $b_{0,1}(x)= \mu \exp(-\mu x)$. Then
\begin{equation}
F(\theta)=
\left(1+\frac{\theta}{\mu-c\lambda/b}\right)^{-\lambda/b}
\end{equation}
Comparing to the solution of the storage model $c=0$ shows
that the additional term leads to an effective decrease of the
parameter $\mu \to \mu - c\lambda/b$. The stationary probability
distribution is given by the gamma-distribution function

$$ \omega_{st}(X)= (\mu-c\lambda/b)^{\lambda/b} X^{\lambda/b-1}
e^{-X(\mu-c\lambda/b)}/\Gamma(\lambda/\beta)\,.$$
The stationary
solution exists if $\mu-c\lambda/b>0$, otherwise the system
undergoes the phase transition with the system overflow likewise
for the model with constant release rate. If $c<1$ there are
two phase transitions (increasing $\lambda$), first of which is
that of the model (\ref{omega:lin}) (Fig.1) and the second is
the system overflow at $1-c\rho/b=0$. If $c>1$ the overflow occurs
earlier than the condition $\lambda=b$ meets and qualitatively the
behaviour of the system is similar to that of constant release
rate (\ref{r:const}), with the transition condition $c\rho/b=1$
instead of former $\rho=a$.

Now let us sketch an example of application of the generalized storage
scheme for the problem of neutron fission process. Set the
generating function $\varphi_{1}(\theta)$ in (\ref{V:gen:exam}) in
the following form

$$\varphi_{1}(\theta)=\lambda_1 \left[1-\sum\limits_{k=0}^{\infty}\pi_{k}\exp(-\theta
k)\right], \quad \sum\limits_{k=0}^{\infty}\pi_k=1 \,,
$$
here $\sum_{k=0}^{\infty}\pi_{k}z^k$ is the generating function of
the neutron number distribution per one elementary fission act
($z=exp(-\theta)$ for a discrete variable); $\pi_k$ are
probabilities of emerging $k$ secondary neutrons (discrete
analogue of the function $b(x)$ in (\ref{phi})); $\lambda_1$ is the
fission intensity (probability of a fission act per time unit),
$\lambda_1=\langle v\rangle\Sigma_f$ with average neutron velocity
$\langle v\rangle$ and macroscopic fission cross section
$\Sigma_f$ \cite{Zweifel}; further, $b=1/l_{ef}$, $l_{ef}$ is the
average neutron lifetime till the absorbtion or escape \cite{Zweifel}.
Let us set $c=1$ and assume that the function $\varphi_0(\theta)$
accounts for the external neutron source with intensity $q\equiv\partial
\varphi_0(\theta)/\partial \theta_{|\theta=0}$ (the smallness
parameter $\gamma$ in (\ref{S:series}) now describes the relation
$\lambda_1/\lambda_0$ of the intensities of fission and external
source events). This probabilistic model is essentially the same
as in the example of the generalized storage scheme (\ref{V:gen:exam})
sketched above. From (\ref{kin:equat:pot}),(\ref{V:gen:exam})

$$\partial \log F(\theta)/\partial t=\left\{-\varphi_{0}(\theta)-[\theta
b -\varphi_{1}(\theta)]\partial /\partial\theta\right\}\log
F(\theta) \, ,$$ that is we arrive at the equation for the
distribution function of the prompt neutrons in the diffusion
single-velocity approximation. The macroscopic equation for the
averages \newline $\langle N\rangle=$ $-\partial \log
F(\theta)/\partial \theta_{|\theta=0}$ is
 $$\frac{d\langle N\rangle}{dt}=\left[\langle v\rangle
 \langle\nu\rangle\Sigma_f-1/l_{ef}\right]\langle N\rangle+q$$
\cite{Zweifel} and coincides with that obtainable apriori from the
stochastic storage model. The neutron reproduction factor is defined as
$k=\lambda_1 \langle\nu\rangle/b=$ $\langle\nu\rangle \langle
v\rangle\Sigma_f l_{ef}$ \cite{Zweifel}, where
$\langle\nu\rangle=$ $\lambda_{1}^{-1}\partial
\varphi_1(\theta)/\partial \theta_{|\theta=0}$ is the average
number of secondary neutrons per one fission act. The expression
for the generalized storage model phase transition
$\lambda_1\langle\nu\rangle/b=1$ corresponds to the reactor
criticality condition $k=1$. Extending probabilistic schemes in
(\ref{V:gen:exam}) beyond the toy model considered here and
introducing vector (multi-comp\-on\-ent) stochastic processes allows
for taking into account the delayed neutrons, as well as various feedbacks
and controlling mechanisms.

\section{Appendix. Derivation of the relations (\ref{stoch:pot}) and (\ref{FDR:2})}
\label{sect:5}

Here we sketch the elementary derivations of the expressions in the text
describing
the kinetic potential of the storage model (expr. (\ref{stoch:pot}))
and that of the stationary FDR through the image of the kinetic
potential (expr.(\ref{FDR:2})).
 The more rigorous and generalizing
derivations can be searched elsewhere, resp.
\cite{Prabhu,Brock,Ryazanov:1993,Stratonovich}.

\subsection{Kinetic potential of the storage model}
From (\ref{storage:eq}) and (\ref{stoch:pot:der}) using the fact that the input rate is
a random process independent on $X(\tau)$ (to simplify notations we set the initial moment $t=0$):

\begin{eqnarray*}
 \frac{1}{\tau} E\left( e^{-\theta A(\tau) + \theta \int_0^\tau r_{\chi}[X(u)] du} -1 \right)
\simeq
\nonumber \\
\frac{1}{\tau}\left\{ E\left(e^{-\theta A(\tau)} \right)
E\left(1+ \theta \int_0^\tau r_{\chi}[X(u)] du  \right) -1 \right\}\,.
\end{eqnarray*}
Then, using (\ref{Levi}) and taking $\tau \to 0$,
$$
V(-\theta,B) = -\varphi(\theta) + \theta \lim_{\tau \to 0} \frac{1}{\tau} \langle
\int_0^{\tau} r_{\chi}[X(u)]du \rangle \,.
$$
The last term of this expression gives $\theta r_{\chi}[X] + O(\tau)$ in the limit
$\tau \to 0$ provided
that the intensity of jumps is {\it finite} which is the case of the considered class of
Poisson processes.

\subsection{Stationary FDR}
We limit ourselves to the nonequilibrium FDR relation only. For
a general case of the detailed balance,
as well as for the generalization to non-Markov processes, the reader is
referred, e.g., to the book \cite{Stratonovich}.

The stationary Fokker-Planck equation is written as

\begin{equation}
{\cal{N}}_{\partial,B} V\left( -\frac{\partial}{\partial B}, B\right)\equiv
\sum\limits_{n=1}^{\infty}\frac{1}{n!}\left( - \frac{\partial}{\partial
B}\right)^n K_n(B)\omega_{st}(B) = 0
\label{app:fpeq}
\end{equation}
Perform over (\ref{app:fpeq}) the operation
$\int \exp(xB)(\cdot) \mathrm{d}B$. Use the relation

$$\int e^{xB} \left( - \frac{\partial}{\partial
B}\right)^n f(B)\mathrm{d}B = x^n \int  e^{xB} f(B) \mathrm{d}B $$
for some $f(B)$ which can be verified with recursive integrations by parts (the terms with
full derivation vanish at $B=\pm \infty$; if the space of states is a semiaxis as in the
storage models, the integration $\int_{0-}^{\infty}$ including the atom in $0$ is assumed).
Then,
\begin{equation}
(\ref{app:fpeq}) \Rightarrow \int e^{xB} \omega_{st}(B) \mathrm{d}B \left(\sum\limits_{n=1}^{\infty}
\frac{1}{n!} x^n K_n(B) \right) \sim R(x,x) =0 \,.
\end{equation}


\begin{thebibliography}{99}

\bibitem{Horst} W. Horsthemke, R. Lefever, {\it Noise-Induced Transitions.
Theory and Applications in Physics, Chemistry, and Biology}
(Springer Verlag, Berlin, 1984)

\bibitem{Prabhu} N.U. Prabhu, {\it Stochastic storage processes }
(Springer Verlag, Berlin, 1980)

\bibitem{Brock}  P.J. Brockwell,  S.I. Resnick, R.L. Tweedie,
Adv. Appl. Prob. {\textbf 14}, 392 (1982)

\bibitem{Risken} H. Risken, {\it The Fokker-Planck Equation} (Springer, Berlin,
1984)

\bibitem{Ryazanov:1990}
V.V. Ryazanov, in {\it Aerosols: Sciense, Indystry, Health and
Environment}, 1 band (Pergamon Press, Kyoto, 1990), p.142;
 V.V. Ryazanov, S.G. Shpyrko, Journal of Aerosol Science, {\textbf 28}, 647 (1997);
 {\it ibid.} {\textbf 28}, 624 (1997)
\bibitem{Ryazanov:1995}
 V.V. Ryazanov, S.G. Shpyrko, in {\it Proc. Int. Conf. on Probabilistic Safety
Assessment Methodology and Applications. PSA'95, Seoul, 1995}
(Atomic Energy Research Institute, Seoul, Korea, 1995), v.1, p.121



\bibitem{Ryazanov:1993}
V.V. Ryazanov, Ukr.Phys.J. {\textbf 38}, 615 (1993)

\bibitem{Ryazanov:2006} V.V.Ryazanov, S.Shpyrko, Cond.Matt.Phys., {\textbf
9}, 71 (2006).

\bibitem{Chechkin:2005:1} A.V.Chechkin {\it et al}, Europhys.Lett. {\textbf
72}, 348 (2005); J.Phys.A:Math.Gen. {\bf 36}, L537 (2003)


\bibitem{Feller:2}  W. Feller, {\it An Introduction to Probability Theory
and its Applications}, vol.2, (John Wiley \& sons, Inc., New York,
1971)

\bibitem{Vankampen}  N.G. Van Kampen, {\it Stochastic Processes in Physics and
Chemistry} (North-Holland Personal Library, 1992)

\bibitem{Gard}  C.W. Gardiner, {\it Handbook of Stochastic Methods for Physics, Chemistry
and the Natural Sciences}, Springer Series in Synergetics, Vol.
13, 3rd edn. (Springer Verlag, Berlin, 2004)

\bibitem{Metzler} R.Metzler,J.Klafter, Phys.Rep.{\bf 339},1 (2000)

\bibitem{Chechkin:2005:2} A.V.Chechkin,V.Yu.Gonchar, J.Klafter, R.Metzler,
Phys.Rev. {\textbf 72 E}, 010101 (R) (2005)

\bibitem{Chechkin:2003} A.V.Chechkin {\it et al}, Phys.Rev.
{\textbf 67 E}, 010102(R) (2003).

\bibitem{Chechkin:2002} A.V.Chechkin, R.Gorenflo, I.M.Sokolov, Phys.Rev. {\textbf 66
E}, 046129 (2002)


\bibitem{Sokolov:2003} I.M.Sokolov, R.Metzler, Phys.Rev. {\textbf
67 E}, 010101(R) (2003).

\bibitem{Uchaikin} V.V.Uchaikin, Uspekhi fizicheskih nauk {\textbf
173}, 847 (2003) [Physics - Uspekhi, {\textbf 46}, 821 (2003)].

\bibitem{Zolotarev} V.M.Zolotarev, {\it One-Dimensional Stable
Distributions}, AMS, Providence, RI, 1986.


\bibitem{Jespersen} S.Jespersen, R.Metzler, and H.S.Fogedby,
Phys.Rev. {\textbf 59 E}, 2736 (1999)

\bibitem{Hongler} M.-O.Hongler, R.Filliger, P.Blanchard, Physica A
{\textbf 370}, 301 (2006)


\bibitem{Stratonovich}  R.L. Stratonovich, {\it Nonlinear Non\-equilibrium
Ther\-mo\-dynamics}, Springer Series in Synergetics, vol. 59
(Springer Verlag, Berlin, 1994)

\bibitem{Zweifel}  P.F. Zweifel, {\it Reactor Physics} (McGraw-Hill, New York,
1973)


\end{thebibliography}
\end{document}